\documentclass[aps,floats]{revtex4}
\usepackage{amsmath,amssymb}
\usepackage{graphicx,epsfig}
\usepackage[greek,english]{babel}
\usepackage{bbold}

\begin{document}
\bibliographystyle {plain}

\pdfoutput=1
\def\oppropto{\mathop{\propto}} 
\def\opsimeq{\mathop{\simeq}}
\def\opoverderline{\mathop{\overline}}
\def\operarrow{\mathop{\longrightarrow}}
\def\opsim{\mathop{\sim}}

\def\fig#1#2{\includegraphics[height=#1]{#2}}
\def\figx#1#2{\includegraphics[width=#1]{#2}}


\title{ Large Deviations for the density and the current  \\
in Non-Equilibrium-Steady-States on disordered rings  } 


\author{ C\'ecile Monthus }
 \affiliation{Institut de Physique Th\'{e}orique, 
Universit\'e Paris Saclay, CNRS, CEA,
91191 Gif-sur-Yvette, France}

\begin{abstract}
The so-called 'Level 2.5' general result for the large deviations of the joint probability of the density and of the currents for Markov Jump processes is applied to the case of $N$ independent particles on a ring with random transition rates. We first focus on the Directed Trap model, where the contractions needed to obtain the large deviations properties of the density alone and of the current alone can be explicitly written in each disordered sample, and where the deformed Markov operator needed to evaluate the generating function of the current can be also explicitly analyzed via its highest eigenvalue and the corresponding left and right eigenvectors. We then turn to the non-directed model, where the tails for large currents $ j \to \pm \infty$ of the rate function for the current alone can still be studied explicitly, either via contraction or via the deformed Markov operator method. We mention the differences with the large deviations properties of the Fokker-Planck dynamics on disordered rings.

\end{abstract}

\maketitle

\section{Introduction}

In the field of Non-Equilibrium Stochastic processes,
 the language of Large Deviations (see the reviews \cite{oono,ellis,review_touchette} and references therein)
has provided an unifying framework, as emphasized in the reviews  \cite{derrida-lecture,harris_Schu,searles,harris,lazarescu_companion,lazarescu_generic}
and in the PhD Theses \cite{vivien_thesis,chetrite_thesis,wynants_thesis}.
However the standard classification of Large Deviations into three levels \cite{oono,review_touchette}, with Level 1 for empirical observables, Level 2 for the empirical measure,
and Level 3 for the empirical process,
has turned out to be insufficient for non-equilibrium phenomena where currents play a major role.
The so-called 'Level 2.5' has thus been introduced to characterize the large deviations properties of the joint distribution of the empirical measure and of the empirical flows. Then the rate function at Level $2.5$ can be 
written as an explicit local functional of the density and of the flows
for various Markovian dynamics,
including Markov Chains (discrete-space and discrete-time) \cite{fortelle_thesis,fortelle_chain,review_touchette},
Markov Jump processes (discrete-space and continuous-time)
\cite{fortelle_jump,wynants_thesis,maes_canonical,maes_onandbeyond,chetrite_formal}
and Diffusions (continuous-space and continuous-time) \cite{wynants_thesis,maes_diffusion,chetrite_formal,engel}.

These large deviations at Level $2.5$ a priori allow to analyze the large deviations 
of the density alone (Level 2), of the flows alone, and more generally of all time-additive observables of the dynamical trajectories,
via the appropriate contractions.
However in practice, the optimization problems required by these contractions can be very difficult to solve explicitly.
Another approach thus consists in analyzing the generating functions of time-additive functionals of the dynamical trajectories
via deformed Markov operators, like the current  (see the reviews \cite{derrida-lecture,lazarescu_companion,lazarescu_generic}
for the case of exclusion processes ) or many other time-additive observables 
of physical interest for the non-equilibrium model under study \cite{lecomte_chaotic,lecomte_thermo,lecomte_formalism,
lecomte_glass,kristina1,kristina2,chetrite_canonical,chetrite_conditioned,touchette_langevin,derrida-conditioned,bertin-conditioned}.

In this paper, we wish to analyze the large deviations properties at Level $2.5$ for the specific case of $N$ independent particles 
moving on a ring of $L$ sites via some Markov Jump process involving random transition rates.
The paper is organized as follows.
In section \ref{sec_general}, we introduce the model and recall the corresponding large deviations framework
with the various approaches.
In section \ref{sec_directed}, we focus on the directed version of the model, 
where many explicit results can be obtained.
In section \ref{sec_nondirected}, we turn to the asymmetric model to analyze the similarities and the differences
with the directed model.
In section \ref{sec_diffusion}, we compare with the case of the Fokker-Planck dynamics on a random ring.
The conclusions are summarized in section \ref{sec_conclusion}.


\section{ Model and large deviations observables }

\label{sec_general}

\subsection{ Master Equation for a single particle on a ring of $L$ sites}

Among the various models for random walks in random media (see the reviews \cite{haus,jpb_review,annphys90,havlin,c_review}),
we focus here on the Markov Jump process on a ring of $L$ sites $x=1,2,..,L$ with periodic boundary conditions ($x+L \equiv x$),
where the probabilities $P_t(x)$ to be at site $x$ at time $t$ evolve with the master equation \cite{derrida}
\begin{eqnarray}
\partial_t P_t(x) = w_{x-1}^+ P_t(x-1) +w_{x+1}^- P_t(x+1)  - (w_x^+ + w_x^-)  P_t(x)
\label{master}
\end{eqnarray}
where the transitions rates $w_x^{\pm}$ from position $x$ towards its neighbors $x \pm 1$ depend on the position $x$
and can be in particular random.
As usual, it is convenient to introduce the ket $ \vert P_t \rangle$ to gather the $L$ components 
$P_t(x) = \langle x \vert P_t \rangle$
and to introduce the matrix
\begin{eqnarray}
 W  =\sum_{x=1}^L \left( w_x^+ \vert x +1 \rangle \langle x \vert 
+ w_x^- \vert x -1 \rangle \langle x \vert - (w_+^+ + w_x^-) \vert x  \rangle \langle x \vert \right)
\label{wmaster}
\end{eqnarray}
 to rewrite the $L$ equations of Eq. \ref{master} as
\begin{eqnarray}
\frac{d}{dt} \vert P_{t} \rangle= W \vert P_t \rangle
\label{masterket}
\end{eqnarray}
The bra $\langle\Omega \vert =\sum_{x=1}^L \langle x \vert$ allows to rewrite the normalization of probabilities at each time $t$ as
\begin{eqnarray}
1 = \sum_{x=1}^L P_t(x)= \langle \Omega \vert P_{t} \rangle
\label{normaomega}
\end{eqnarray}
The vanishing of its time derivative 
\begin{eqnarray}
0 = \frac{d}{dt}  \langle \Omega \vert P_{t} \rangle = \langle \Omega \vert W \vert P_{t} \rangle
\label{normaomegaderi}
\end{eqnarray}
yields that $\langle\Omega \vert = \langle L_0 \vert$ is the Perron-Frobenius 
Left eigenvector associated to the highest eigenvalue $E_0=0$ of the operator $W$,
while the corresponding Perron-Frobenius  Right eigenvector $\vert R_0 \rangle = \vert \rho_{st} \rangle$ corresponds to the stationary state.
The relaxation towards this stationary state is governed by the other $(L-1)$ eigenvalues $E_n<0$ via the spectral 
spectral decomposition
\begin{eqnarray}
W = \sum_{n=0}^{L-1} E_n \vert R_n \rangle \langle L_n \vert
\label{srpectral}
\end{eqnarray}
that leads to
\begin{eqnarray}
\vert P_{t} \rangle=e^ {W t}  \vert  P_{0} \rangle =   \sum_{n=0}^{L-1} e^{ E_n t }  \vert R_n \rangle \langle L_n \vert  P_{t} \rangle
=  \vert \rho_{st} \rangle + \sum_{n=1}^{L-1} e^{ E_n t }  \vert R_n \rangle \langle L_n \vert  P_{0} \rangle
\label{relax}
\end{eqnarray}
The steady state corresponds to the constant current $j_{st}$ along the ring $x=1,2,..,L$
\begin{eqnarray}
 j_{st} = \rho_{st}(x) w_x^+ -  \rho_{st}(x+1) w_{x+1}^-
\label{jstconserv}
\end{eqnarray}
while the density is normalized  $1=\sum_{x} \rho_{st}(x) $.
The explicit solution in terms of all the rates reads \cite{derrida}
\begin{eqnarray}
\rho_{st}(x) 
&& =j_{st} 
 \frac{  \frac{1}{w_x^+} \left[ 1+\displaystyle \sum_{n=1}^{L-1}  \prod_{y=1}^n \frac{w_{x+y}^-  }{w_{x+y}^+} \right]  }
{ 1- \displaystyle  \prod_{z=1}^L \frac{w_z^-  }{w_z^+}  }
\label{rhost}
\end{eqnarray}
with the stationary current
\begin{eqnarray}
j_{st} =  \frac{ 1- \displaystyle    \prod_{z=1}^L \frac{w_{z}^-  }{w_{z}^+}  }
{\displaystyle  \sum_{x=1}^L \frac{1}{w_x^+} \left[ 1+ \sum_{n=1}^{L-1}  \prod_{y=1}^n \frac{w_{x+y}^-  }{w_{x+y}^+} \right]  }
\label{jst}
\end{eqnarray}
The properties of this steady state as a function of the statistics of the random transition rates
is discussed in detail in \cite{derrida}.
In the following, we focus instead on the large deviations far from this steady state in a given disordered ring.

\subsection{ Reminder on the large deviations for the empirical density $\rho(x)$ and the empirical current $j$  }

Following \cite{maes_onandbeyond}, we wish to consider 
the dynamics of $N$ independent particles on the ring,
i.e. $N$ independent jump Markov processes with trajectories $x^{(k)}(t)$.
The dynamical fluctuations can be analyzed \cite{maes_onandbeyond} 
via the ensemble-empirical-density at site $x$ at time $t$
\begin{eqnarray}
\rho_t(x) = \frac{1}{N} \sum_{k=1}^N \delta \left( x^{(k)}(t) -x \right)
\label{ensembleempirical}
\end{eqnarray}
and the ensemble-empirical-jump-density from site $x$ to site $y$ at time $t$
\begin{eqnarray}
q_t(y,x) = \frac{1}{N} \sum_{k=1}^N \delta \left( x^{(k)}(t^+) -y \right) \delta \left( x^{(k)}(t^-) -x \right)
\label{ensemblejump}
\end{eqnarray}
The antisymmetric part corresponds to the ensemble-empirical-currents from site $x$ to site $y$ at time $t$
\begin{eqnarray}
j_t(y,x) =q_t(y,x)  - q_t(x,y) = - j_t(x,y)
\label{ensemblecurrent}
\end{eqnarray}
while the symmetric part usually called 'activity' or 'traffic' 
\begin{eqnarray}
a_t(y,x) =q_t(y,x)  + q_t(x,y) = a_t(x,y)
\label{ensembleactivity}
\end{eqnarray}
is of course also interesting \cite{wynants_thesis,maes_canonical,maes_onandbeyond}
but will not be considered further here.

For the ring model where the jumps from $x$ occur only towards $y=(x \pm 1)$, it is thus convenient to
introduce the following simplified notation for the current on each link $(x+1 \leftarrow x)$
\begin{eqnarray}
j_t(x) \equiv j_t(x+1,x) =q_t(x+1,x)  - q_t(x,x+1) = - j_t(x,x+1)
\label{defjtx}
\end{eqnarray}
Dropping the boundary term coming from the initial condition (see \cite{maes_onandbeyond} if you wish to keep it),
 the general formula of Ref. \cite{maes_onandbeyond}
yields for the present ring model that it is possible to observe
the empirical density $\rho_t(x)$ and the empirical currents $j_t(x)$ during some time interval $0 \leq t \leq T$
only if they satisfy the consistency constraint required by the conservation of probability
\begin{eqnarray}
\partial_t \rho_t(x) = j_t(x-1)-j_t(x)
\label{tconstraint}
\end{eqnarray}
and that the corresponding probability then
follows the large deviation form for large $N$
\begin{eqnarray}
 {\cal P} \left [ \rho_{0 \leq t \leq T} (.) , j_{0 \leq t \leq T}(.) \right] 
\opsimeq_{N \to +\infty}  e^{ - N{\cal I}  \left[  \rho_{0 \leq t \leq T} (.) , j_{0 \leq t \leq T}(.) \right]    }
\label{lddyna}
\end{eqnarray}
with the time-dependent rate function \cite{maes_onandbeyond}
\begin{eqnarray}
 {\cal I}  \left[ \rho_{0 \leq t \leq T} (.) , j_{0 \leq t \leq T}(.)  \right]  = \int_0^T dt \sum_{x=1}^L {\cal L} (j_t(x) ,\rho_t(x),\rho_t(x+1))
\label{ratedyna}
\end{eqnarray}
where the contribution of the link $(x,x+1)$ at time $t$ only depends on the current $j_t(x)$ on this link
and on the two densities $\rho_t(x) $ and $\rho_t(x+1) $ at the boundaries of this link, and reads \cite{maes_onandbeyond}
\begin{eqnarray}
  {\cal L} (j_t(x) ,\rho_t(x),\rho_t(x+1))
&& =   j_t(x) \ln \left( \frac{\sqrt{ j_t^2(x) + 4 \rho_t(x) w_x^+ \rho_t(x+1) w_{x+1}^-  }  + j_t(x) }
{ 2\rho_t(x) w_x^+}  \right)  
 \nonumber \\ &&
 -\sqrt{ j_t^2(x) + 4 \rho_t(x) w_x^+ \rho_t(x+1) w_{x+1}^-  }  + \rho_t(x) w_x^+ + \rho_t(x+1) w_{x+1}^-   
\label{ratebondtime}
\end{eqnarray}

Since the dynamical constraint of Eq. \ref{tconstraint} is difficult to take into account in practice
when one wishes to analyze all possible dynamical fluctuations,
we will only consider the empirical-time-averages of the density and the current \cite{maes_onandbeyond}
\begin{eqnarray}
\rho(x) && \equiv \frac{1}{T} \int_0^T dt \rho_t(x) 
\nonumber \\
j(x) && \equiv \frac{1}{T} \int_0^T dt j_t(x) 
\label{timeempirical}
\end{eqnarray}
Then it is possible to observe the empirical density $\rho(x)$ and the empirical currents $j(x)$ 
only if they satisfy the stationary version of the consistency constraint of Eq. \ref{tconstraint}
\begin{eqnarray}
 j(x-1)=j(x) = j
\label{divj}
\end{eqnarray}
i.e. the current $j(x)$ has to take the same value $j$ on each link $x=1,2,..,L$ along the ring
(in higher dimensions, the constraint is that the discrete divergence of the current should vanish \cite{maes_onandbeyond}).
Then the corresponding probability to observe the empirical density $\rho(x)$ normalized to unity
\begin{eqnarray}
1=\sum_{x=1}^L \rho(x) 
\label{normarho}
\end{eqnarray}
and the current $j$
follows the large deviation form directly inherited from Eq. \ref{lddyna}
\begin{eqnarray}
 {\cal P} \left[ \rho(.) , j \right]  \opsimeq_{N \to +\infty}  e^{ - N T  I \left[ \rho(.) , j  \right]   }
\label{proba2.5}
\end{eqnarray}
with the time-independent rate function \cite{maes_onandbeyond})
\begin{eqnarray}
 I \left[ \rho(.) , j   \right]  =  \sum_{x=1}^L 
&& [   j \ln \left( \frac{\sqrt{ j^2 + 4 \rho(x) w_x^+ \rho(x+1) w_{x+1}^-  }  + j}
{ 2\rho(x) w_x^+}  \right)  
 \nonumber \\ &&
 -\sqrt{ j^2 + 4 \rho(x) w_x^+ \rho(x+1) w_{x+1}^-  }  + \rho(x) w_x^+ + \rho(x+1) w_{x+1}^-   ]
\label{rate2.5masterrhojd1}
\end{eqnarray}
As the exponential form in $(NT)$ of Eq. \ref{proba2.5} suggests, this stationnary large deviation result
holds also for a single particle $N=1$ in the large time limit $T \to +\infty$ and
the time-independent rate function of Eq. \ref{rate2.5masterrhojd1} is usually derived
within this framework \cite{fortelle_jump,wynants_thesis,maes_canonical,chetrite_formal}.
The alternative point of view of Ref. \cite{maes_onandbeyond} that we have summarized here
enables to relate the usual time-independent rate function of Eq. \ref{rate2.5masterrhojd1}
to its time-dependent counterpart of Eqs \ref{ratedyna} and \ref{ratebondtime}
and allows for generalizations to open systems where the total number of particles is not conserved \cite{c_largedevopen}
and to interacting particles \cite{c_lardevinter}.

From the joint rate function $ I \left[ \rho(.) , j   \right]  $ of Eq. \ref{rate2.5masterrhojd1},
the rate function $ I_{density} [\rho(.)]$ for the density $\rho(x)$ alone and 
the rate function $ I_{current} ( j ) $ for the current $j$ alone
can be then obtained via contractions, as discussed in the next sections,
but before it is useful to recall the alternative approach based on generating functions.

\subsection{ Generating functions via the deformed Markov operator approach  }

Another point of view consists in considering the corresponding generating function
via the introduction of generalized chemical potentials $\mu(x)$ and $\nu$
associated to the empirical density $\rho(x)$ and to the empirical current $j$ on each bond along the ring (Eq. \ref{divj})
\begin{eqnarray}
 Z\left[\mu(.),\nu\right]  && \equiv 
<  \ \ e^{ \displaystyle N T \left( \sum_{x=1}^L \mu(x) \rho(x)  +  \nu j \right) } > 
\label{z2.5}
\end{eqnarray}

On one hand, it can be evaluated from the joint probability of Eq. \ref{proba2.5} 
for the empirical density density $\rho(x)$ and the empirical current $j$ 
\begin{eqnarray}
 Z\left[\mu(.),\nu\right]  = \int d\rho(1) ... \int \rho(L)  \int dj \ 
e^{ \displaystyle N T  \left( \sum_{x=1}^L \mu(x) \rho(x)  +  \nu j    -  I \left[ \rho(.) , j  \right] \right)  }
\oppropto_{NT \to +\infty} e^{ \displaystyle N T E[\mu(.),\nu] }
\label{gene2.5}
\end{eqnarray}
where $E[\mu(.),\nu]$ obtained from via the saddle-point evaluation 
\begin{eqnarray}
  E[\mu(.),\nu ] = \max \limits_{\rho(.), j  }
\left[ \sum_{x=1}^L \mu(x) \rho(x)  +  \nu j    -  I \left[ \rho(.) , j  \right]   \right]
\label{legendreEI}
\end{eqnarray}
corresponds to the multidimensional Legendre transform of the rate function $I \left[ \rho(.) , j  \right]  $ 
and thus contains the same information.

On the other hand, the generating function of Eq. \ref{z2.5} can be rewritten 
in terms of the 
 deformed Markov operator with respect to the initial operator $W$ of Eq. \ref{wmaster}
\begin{eqnarray}
 W^{[\mu(.),\nu ]}  =
\sum_{x=1}^L \left( w_x^+ e^{\frac{\nu}{L}}  \vert x +1 \rangle \langle x \vert
 + w_x^- e^{-\frac{\nu}{L}}  \vert x -1 \rangle \langle x \vert
 - (w_x^+ + w_x^- -\mu(x) ) \vert x  \rangle \rangle x \vert \right)
\label{wmasterdeformed}
\end{eqnarray}
and the rate function $E[\mu(.),\nu] $ 
corresponds to the highest eigenvalue of this deformed Markov operator.
In practice, this method is usually applied to study the generating function of the current alone,
i.e. when the chemical potentials $\mu(x)$ associated to the density $\rho(x)$ vanish $\mu(x)=0$
\begin{eqnarray}
 Z\left[\nu\right]  && \equiv 
<  \ \ e^{  N T  \nu j  } > \oppropto_{NT \to +\infty} e^{  N T E(\nu) }
\label{znu}
\end{eqnarray}
where Eq. \ref{legendreEI} becomes the Legendre transform for the single pair $(j,\nu)$ of variables
\begin{eqnarray}
  E(\nu)= \max \limits_{j  }
\left[   \nu j    -  I_{current} ( j )   \right]
\label{legendrejnu}
\end{eqnarray}

This approach based on deformed Markov operators has been used in particular
for the generating function of the current in interacting models like exclusion processes
 (see the reviews \cite{derrida-lecture,lazarescu_companion,lazarescu_generic}
and references therein) 
as well as for many other generating function of time-additive functionals of dynamical trajectories
of various non-equilibrium models \cite{lecomte_chaotic,lecomte_thermo,lecomte_formalism,
lecomte_glass,kristina1,kristina2,chetrite_canonical,chetrite_conditioned,touchette_langevin}.


\section{ Application to the DIRECTED TRAP MODEL on a RING }

\label{sec_directed}

In this section, we focus on the Directed Trap Model \cite{jpdir,aslangul,comptejpb,directedtrapandsinai,vanwijland_trap}
on the ring geometry,
that corresponds to the special case of the master Eq \ref{master}
where all the transition rates in the negative direction vanish along the ring
\begin{eqnarray}
w_x^- =0
\label{wmoinszero}
\end{eqnarray}
while the transition rates in the positive direction are interpreted as the inverses of trapping times $\tau_x$
\begin{eqnarray}
w_x^+  = \frac{1}{\tau_x}
\label{trapping}
\end{eqnarray}
The master Eq \ref{master} simplifies into
\begin{eqnarray}
\partial_t P_{t}(x) = \frac{P_t(x-1)}{\tau_{x-1} }- \frac{P_t(x)}{\tau_{x} }
\label{masterdirected}
\end{eqnarray}
The steady state of Eq. \ref{rhost} and the stationary current of Eq. \ref{jst} then reduce to
\begin{eqnarray}
\rho_{st}(x) && =  j_{st} \tau_x
\nonumber \\
j_{st} && =  \frac{ 1  } {\displaystyle  \sum_{x=1}^L \tau_x  }
\label{jstdirected}
\end{eqnarray}
so that the physical meaning is completely obvious : 
the density $\rho_{st}(x)$ is simply proportional to the trapping time $\tau_x$,
while the current $j_{st}$ is the inverse of the sum of the $L$ trapping times along the ring.
When the trapping times $\tau_x$ are  independent random variables
distributed with the power-law distribution 
\begin{eqnarray} 
L_{\mu}(\tau) =\frac{\mu }{ \tau^{1+\mu}} \theta( \tau \geq 1)
\label{lawtrap}
\end{eqnarray}
the region $0<\mu<1$, where the averaged trapping time $\overline{\tau} = \int_0^{+\infty} d \tau \tau L_{\mu}(\tau) = \infty$ diverges,
corresponds to the anomalous diffusion phase $x(t) \propto t^{\mu}$ for the process defined on the infinite line and 
 has been studied from various points of views \cite{jpdir,aslangul,comptejpb,directedtrapandsinai,vanwijland_trap}.
More generally, various other trap models have been also analyzed in relation with anomalously slow glassy behaviors 
\cite{jpb_weak,dean,jp_pheno,bertinjp1,bertinjp2,trapsymmetric,trapnonlinear,trapreponse,trap_traj}.
In the present ring geometry, the anomalous behavior of the stationary current of Eq. \ref{jstdirected} in the region $0<\mu<1$
directly comes from the behavior of the L\'evy sum $\sum_{x=1}^L \tau_x \propto  L^{\frac{1}{\mu}}$ dominated by the largest trapping time
(see the review \cite{jpb_review}).

In the following, we focus on the large deviations properties far from the steady state of Eq. \ref{jstdirected}
within a given disordered sample characterized by a fixed sequence of trapping times.

\subsection{ Large deviation for the joint probability of density $\rho(x)$ and the current $j$ }

The large deviation rate function of Eq. \ref{rate2.5masterrhojd1}
for the joint distribution of the density $\rho(x)$ and the positive current $j \geq 0$
 (as a consequence of the directed character of the model)
reduces to
\begin{eqnarray}
 I^{Directed}[\rho(.) ;j] 
&& = \sum_{x=1}^L  \left[   j \ln \left( \frac{ j  \tau_x } { \rho(x) }  \right)  -j + \frac{\rho(x) }{\tau_x }   \right]
\label{rate2.5directed}
\end{eqnarray}
At zero current $j=0$, it reads
\begin{eqnarray}
 I^{Directed}[\rho(.) ;j=0] 
&& = \sum_{x=1}^L  \frac{\rho(x) }{\tau_x } 
\label{rate2.5directedzero}
\end{eqnarray}
while the decay for large current $j \to +\infty$ is given by the expansion
\begin{eqnarray}
 I^{Directed}[\rho(.) ;j] 
&& =L  j \ln j +  j \sum_{x=1}^L  \left[   \ln \left( \frac{   \tau_x } { \rho(x) }  \right)  - 1 \right] + \sum_{x=1}^L \frac{\rho(x) }{\tau_x }   \opsimeq_{j \to +\infty} L j \ln j
\label{rate2.5directedinfinity}
\end{eqnarray}

\subsection{ Large deviations for the density $\rho(x)$ alone  }

The optimization with respect to the current $j$ in Eq. \ref{rate2.5directed}
\begin{eqnarray}
0 && = \frac{d}{dj}  I^{Directed}[\rho(x) ;j] = \sum_{x=1}^L   \ln \left( \frac{ j  \tau_x } { \rho(x) }  \right) 
   = L \ln j -  \sum_{x=1}^L  \ln \left( \frac{\rho(x) }{\tau_x }  \right)
\label{joptopti}
\end{eqnarray}
yields the optimal current $ j_{opt}   $ as a function of the given density $\rho(x)$
\begin{eqnarray}
  j_{opt}  = e^{\displaystyle  \frac{1}{L}  \sum_{x=1}^{L}   \ln \left( \frac{\rho(x) }{\tau_x }  \right) }
 =  \left[ \prod_{x=1}^L \frac{\rho(x) }{\tau_x }  \right]^{\frac{1}{L} }
\label{joptdirected}
\end{eqnarray}
Plugging this optimal value into Eq. \ref{rate2.5directed} 
yields the rate function for the density $\rho(x)$ alone
\begin{eqnarray}
 I^{Directed}_{density}[\rho(.) ] && = I^{Directed}_{2.5}[\rho(x) ;j_{opt} ] 
= \sum_{x=1}^L \left[ \frac{\rho(x) }{\tau_x }     \right] - L j_{opt} 
\nonumber \\
&&   =  \sum_{x=1}^L\left( \frac{\rho(x) }{\tau_x }    \right) -L    e^{\displaystyle  \frac{1}{L}  \sum_{x=1}^{L}   \ln \left( \frac{\rho(x) }{\tau_x }  \right) }
   =  \sum_{x=1}^L\left( \frac{\rho(x) }{\tau_x }    \right) -L    \left[ \prod_{x=1}^L \frac{\rho(x) }{\tau_x }  \right]^{\frac{1}{L} }
\label{i1directed}
\end{eqnarray}
This simple example show explicitly how the contraction over the current $j$ transforms 
the additive local functional $ I^{Directed}[\rho(x) ;j]$  of Eq. \ref{rate2.5directed}
into a non-additive functional $ I^{Directed}_{density}[\rho(.) ]  $ for the density alone.
Of course this phenomenon is completely general and explains why in Non-Equilibrium-Steady-States,
the large deviations properties of the density alone cannot be described by additive local functional
as a consequence of the currents flowing through the whole sample and that introduce density-correlations.

In order to see more clearly the physical meaning of the optimal current $j_{opt}$ as a function of the imposed density $\rho(x)$,
we may use the steady state of Eq. \ref{jstdirected} to replace $\tau_x = \frac{\rho_{st}(x)}{j_{st} } $
 into  Eq. \ref{joptdirected}
to obtain
\begin{eqnarray}
  j_{opt}  = j_{st} \ e^{\displaystyle  \frac{1}{L}  \sum_{x=1}^{L}   \ln \left( \frac{\rho(x) }{ \rho_{st}(x) }  \right) }
\label{joptdirectedst}
\end{eqnarray}
As a consequence, $ j_{opt} $ will be much smaller than $j_{st}$ if the density $\rho(x)$ happens
to be much smaller than $\rho_{st}(x)$ on most sites along the ring, i.e. when the normalization is concentrated on a few sites,
while $ j_{opt} $ will be much bigger than $j_{st}$ if the density happens to be nearly uniform along the ring $\rho(x) \sim 1/L$.
We will recover the same idea in the next section concerning the opposite contraction.

\subsection{ Large deviations for the current $j$ alone }

The optimisation of the rate function of Eq. \ref{rate2.5directed}
with respect to the density $\rho(x)$ submitted to the normalisation condition of Eq. \ref{normarho}
that can be taken into account via the Lagrange multiplier $k(j)$ that will depend on $j$
yields
\begin{eqnarray}
0 && = \frac{ \partial}{ \partial \rho(x) } \left[   I^{Directed}[\rho(.) ;j] + k(j) \left( \sum_x \rho(x)-1 \right)  \right]
  = - \frac{ j  } { \rho(x) }  + \frac{1}{\tau_x }   +k(j)
\label{rate2.5masterrhojd1derirho}
\end{eqnarray}
The optimal density for a given current $j$ thus reads
\begin{eqnarray}
 \rho_{opt}(x)  && = \frac{j }{ \frac{1}{\tau_x }   +k(j) } =  \frac{j \tau_x}{ 1   +\tau_x k(j) } 
\label{jstconservpho}
\end{eqnarray}
where the Lagrange multiplier $k(j)$ has to be chosen to ensure the normalization
\begin{eqnarray}
 1=\sum_{x=1}^L \rho_{opt}(x)  && = j \sum_{x=1}^L     \frac{ \tau_x}{ 1   + \tau_x k(j)  } 
\label{1kjdirected}
\end{eqnarray}
Plugging this optimal solution into Eq. \ref{rate2.5directed} yields the rate function for the current $j$ alone
\begin{eqnarray}
 I^{Directed}_{current} [j] && = I^{Directed}[\rho_{opt}(.) ;j] 
= \sum_{x=1}^L  \left[   j \ln \left( \frac{ j  \tau_x } { \rho_{opt}(x) }  \right)  -j + \frac{\rho_{opt}(x) }{\tau_x }   \right]
\nonumber \\
&& 
= j \sum_{x=1}^{L} \ln \left(  1+ \tau_x k(j)  \right) -k(j)
\label{ijdirectedimp}
\end{eqnarray}
This form is however somewhat implicit since the Lagrange multiplier $k(j)$ is defined via Eq.  \ref{1kjdirected}.
One possibility consists in writing the solution parametrically,
by considering instead that the current $j$ is a function of the Lagrange multiplier $k$ via Eq. \ref{1kjdirected}
\begin{eqnarray}
 j(k) = \frac{ 1}{ \displaystyle \sum_{x=1}^L     \frac{ \tau_x}{ 1   + \tau_x k  }  }
\label{1kjdirectedinverse}
\end{eqnarray}
while the rate function of Eq. \ref{ijdirectedimp} is also written as a function of $k$
\begin{eqnarray}
 I^{Directed}_{current} [j(k)] && 
=  \frac{ \displaystyle\sum_{x=1}^{L} \ln \left(  1+ \tau_x k \right) }{ \displaystyle \sum_{x=1}^L     \frac{ \tau_x}{ 1   + \tau_x k  }  }   
-k
\label{ijdirected}
\end{eqnarray}

In this language, $k=0$ corresponds to the stationary current $j(k=0)=j_{st}$ of Eq. \ref{jstdirected}
where the rate function vanishes $ I^{Directed}_{current} [j(k=0)] =0 $ as it should.

The limit $k \to +\infty$ corresponds to the limit of large current $j \to +\infty$
\begin{eqnarray}
 j(k) &&  \opsimeq_{k \to +\infty}  \frac{ k }{ L }
\nonumber \\
 I^{Directed}_{current} [j(k)] && 
\opsimeq_{k \to +\infty}  
  \frac{k}{L} \sum_{x=1}^{L} \ln \left(  \tau_x k \right) -k
\label{kinftydirected}
\end{eqnarray}
leading to
\begin{eqnarray}
 I^{Directed}_{current} [j] && 
\opsimeq_{j \to +\infty}  
  j \sum_{x=1}^{L} \ln \left( j L  \tau_x  \right) - j L 
= L  j \ln j + j  \sum_{x=1}^{L} \ln \left(  L  \tau_x  \right) - j L 
\label{ijinftydirected}
\end{eqnarray}
where the factor $L$ in the logarithm comes from the uniform optimal density at large current,
where Eq. \ref{jstconservpho} becomes $\rho_{opt}(x)  \simeq \frac{j }{ k(j) } \simeq \frac{1}{L}$,
so that the comparison with the case of large current at fixed density of Eq. \ref{rate2.5directedinfinity} is clear.

The opposite boundary of vanishing current $j=0$ is 
reached in Eq. \ref{1kjdirectedinverse}
for the singular value 
 \begin{eqnarray}
k(j=0) =   -  \frac{ 1 }{ \left(  \max \limits_{1 \leq x \leq L } \tau_x \right) }
\label{kj0}
\end{eqnarray}
determined by the largest trapping time in the ring, and the corresponding rate function 
of Eq. \ref{ijdirected} reduces to
\begin{eqnarray}
 I^{Directed}_{current} [j=0] && =  -k(j=0) = \frac{ 1 }{ \left(  \max \limits_{1 \leq x \leq L } \tau_x \right) }
\label{ijdirectedzerojsing}
\end{eqnarray}
The comparison with Eq. \ref{rate2.5directedzero} shows that the corresponding optimal density $\rho_{opt}(x)$
is a delta function of the site $x_{max}$ having the maximal trapping time $\tau_{max}$.

In summary, with respect to the steady state $(\rho_{st},j_{st})$ of Eq. \ref{jstdirected} determined by the trapping times,
the region of larger currents $j> j_{st}$ corresponds to more homogeneous optimal densities $\rho_{opt}$ than $\rho_{st} $,
up to the full homogeneity $\rho_{opt}(x)  \simeq \frac{1}{L}$ reached in the limit $j \to +\infty$,
while the region of smaller currents $j< j_{st}$ corresponds to more inhomogeneous optimal densities $\rho_{opt}$ than $\rho_{st} $,
up to the most extreme inhomogeneity $\rho_{opt}(x)  \simeq \delta_{x,x_{max}}$ reached in the limit of vanishing current $j =0$

It is now interesting to compare with the approach based on the generating function of the current.

\subsection{ Generating function $Z(\nu)$ of the current $j$ via the deformed Markov operator $W^{\nu}$  }

For the Directed Trap model, 
the deformed operator of Eq. \ref{wmasterdeformed} for the current alone (no chemical potential for the density)
reduces to
\begin{eqnarray}
 W^{[\nu ]}  =
\sum_{x=1}^L \left(\frac{ e^{\frac{\nu}{L}} }{\tau_x } \vert x +1 \rangle \langle x \vert 
 -  \frac{1} {\tau_x } \vert x  \rangle \langle x \vert \right)
\label{wmasterdeformednudirected}
\end{eqnarray}
This deformed operator for the trap model has been already studied in
 \cite{vanwijland_trap} with the translation of notation $\nu=-s L$,
where the parameter $s$ 
 is conjugated to the total number of jumps in order to characterize the glassy character of the dynamics
\cite{vanwijland_trap}.
It is nevertheless interesting to mention here some of its properties
in the present language to make the link with the previous section,
and to compare with the non-directed model of the next section.

 One wishes to compute the highest eigenvalue $E (\nu)$ with the corresponding right eigenvector $ \vert R^{\nu}\rangle$ 
and left eigenvector $\langle L^{\nu} \vert$
\begin{eqnarray}
 W^{[\nu ]} \vert R^{\nu}\rangle && =E (\nu)\vert R^{\nu}\rangle
\nonumber \\
\langle L^{\nu} \vert W^{[\nu ]}  && =\langle L^{\nu} \vert E (\nu)
\label{wmastereigen}
\end{eqnarray}
Their components satisfy simple recurrences and can be thus evaluated in terms of the first components
 \begin{eqnarray}
R^{\nu}(x) = \frac{   \frac{ e^{\frac{\nu}{L}} }{\tau_{x-1} }      }{ \frac{1}{\tau_x} + E(\nu)  }   R^{\nu}(x-1) 
= \left[  \prod_{y=2}^x  \frac{   \frac{ e^{\frac{\nu}{L}} }{\tau_{y-1} }      }{ \frac{1}{\tau_y} + E(\nu)  }       \right]R^{\nu}(1) 
\label{righteigen}
\end{eqnarray}
and
 \begin{eqnarray}
L^{\nu}(x) =\frac{ 1+ E(\nu) \tau_{x-1} }{ e^{\frac{\nu}{L}}  }  L^{\nu}(x-1) 
= \left[  \prod_{y=2}^x  \frac{ 1+ E(\nu) \tau_{y-1} }{ e^{\frac{\nu}{L}}  }    \right]     L^{\nu}(1) 
\label{lefteigen}
\end{eqnarray}
while the periodic boundary conditions $R^{\nu}(L+1)=R^{\nu}(1)  $ and $L^{\nu}(L+1)=L^{\nu}(1)   $
yields that the eigenvalue $E(\nu)$ should satisfy
 \begin{eqnarray}
e^{\nu} =  \prod_{x=1}^L \left( 1+ E(\nu) \tau_{x}  \right) 
\label{enueigen}
\end{eqnarray}
and $(1+ E(\nu) \tau_{x} ) \geq 0$ for all $x=1,2,..,L$ in order to ensure the positivity of the components 
of the Perron-Frobenius eigenvectors $R^{\nu}(x)  $ and $L^{\nu}(x) $ 
 \begin{eqnarray}
 E(\nu)  \geq \max \limits_{1 \leq x \leq L } \left(  - \frac{ 1 }{\tau_x} \right) = -  \frac{ 1 }{ \left(  \max \limits_{1 \leq x \leq L } \tau_x \right) }
\label{enudomain}
\end{eqnarray}
This bound is relevant only in the region $\nu<0$ where the energy is negative $E(\nu)<0$,
while in the region $\nu>0$, the energy is positive $E(\nu)>0$. 
Instead of the function $E(\nu)$, it is simpler to consider its inverse 
 \begin{eqnarray}
\nu(E)  =  \sum_{x=1}^L \ln \left( 1+ E \tau_{x}  \right) 
\label{nue}
\end{eqnarray}
in each given disordered sample.

In this language, the cumulants of the current around its stationary value $j_{st}$ 
are related to the behavior near the origin $\nu = 0$, where the two sides
 $E>0$ and $E<0$ have different properties as discussed in detail in \cite{vanwijland_trap}.
Here we will thus instead focus on the tail for large current $j \to +\infty$
and on the other boundary at zero current $j=0$.

The information about large currents $ j \to +\infty$ 
is contained in the region $\nu \to +\infty$, where the energy is also large $E(\nu) \to +\infty$
and follows the asymptotic behavior (Eq \ref{enueigen} )
\begin{eqnarray}
E(\nu) \opsimeq_{\nu \to +\infty} e^{ \displaystyle \frac{ \nu}{L}  -   \frac{1}{L} \sum_{x=1}^L \ln \left(   \tau_{x}  \right)    }
\label{nuelarge}
\end{eqnarray}
that corresponds to the Legendre transform (Eq. \ref{legendrejnu}) of Eq. \ref{ijinftydirected} as it should.

The information about zero current $j =0$ 
is contained in the opposite limit $\nu \to -\infty$ where the energy reaches its minimal value of Eq. \ref{enudomain}
 \begin{eqnarray}
 E(\nu)  \opsimeq_{\nu \to - \infty}  -  \frac{ 1 }{ \left(  \max \limits_{1 \leq x \leq L } \tau_x \right) }
\label{enulimit}
\end{eqnarray}
that directly represents the opposite of Eq. \ref{ijdirectedzerojsing} via the Legendre transform (Eq. \ref{legendrejnu}) as it should.


\section{ Application to the non-directed disordered ring }

\label{sec_nondirected}

In this section, we return to the non-directed dynamics of Eq. \ref{master}
in order to analyze its large deviations with respect to the steady state of Eq. \ref{rhost} and Eq \ref{jst}

\subsection{ Large deviations for the empirical density $\rho(x)$ and the empirical current $j$}

Let us recall the rate function of Eq. \ref{rate2.5masterrhojd1}
for the joint distribution of the density $\rho(x)$ and of the current 
\begin{eqnarray}
 I \left[ \rho(.) , j   \right]  =  \sum_{x=1}^L 
&& [   j \ln \left( \frac{\sqrt{ j^2 + 4 \rho(x) w_x^+ \rho(x+1) w_{x+1}^-  }  + j}
{ 2\rho(x) w_x^+}  \right)  
 \nonumber \\ &&
 -\sqrt{ j^2 + 4 \rho(x) w_x^+ \rho(x+1) w_{x+1}^-  }  + \rho(x) w_x^+ + \rho(x+1) w_{x+1}^-   ]
\label{rate2.5}
\end{eqnarray}
to stress the similarities and the differences with the Directed Trap model analyzed in the previous section.

The first essential difference is of course that the current varies now on the whole interval $j \in ]-\infty,+\infty[$ (instead of $j \in [0,+\infty[$ for
the directed model), so that 
the rate function displays the Gallavotti-Cohen symmetry 
\cite{derrida-lecture,harris_Schu,searles,harris,chetrite_thesis,lazarescu_companion,lazarescu_generic}
between two opposite values of the current $(\pm j)$ for each fixed density $\rho(x)$ 
\begin{eqnarray}
 I_{2.5}[\rho(x) ;j] -  I_{2.5}[\rho(x) ;- j] 
&& =j  \sum_{x=1}^{L}  \ln \left( \frac{ \rho(x+1) w_{x+1}^- } { \rho(x) w_x^+}  \right) 
= j \sum_{x=1}^{L} \left(  \ln w_x^-  - \ln w_x^+  \right) = j \ln \left(  \prod_{x=1}^L \frac{w_x^-}{w_x^+} \right)
\label{symjmj}
\end{eqnarray}
where the factor $\ln \left(  \prod_{x=1}^L \frac{w_x^-}{w_x^+} \right)  $ directly measures the irreversibility of the dynamics along the ring.

The special value for zero current $j=0$ reads
\begin{eqnarray}
I [\rho(x) ;j=0] 
&& = \sum_{x=1}^{L}
 [   -2 \sqrt{  \rho(x) w_x^+ \rho(x+1) w_{x+1}^-  }  + \rho(x) w_x^+ + \rho(x+1) w_{x+1}^-   ]
 \nonumber \\ &&
 =  \sum_{x=1}^{L} \left(   \sqrt{  \rho(x) w_x^+} - \sqrt{ \rho(x+1) w_{x+1}^-  }   \right)^2
\label{rate2.5masterrhojd1jzero}
\end{eqnarray}
while the tails for large currents $j \to - \infty$ and $j \to +\infty$ are given by
\begin{eqnarray}
 I[\rho(x) ;j] 
&& \opsimeq_{j \to - \infty} 
L \vert j \vert \ln \vert j \vert  - \vert j \vert \left[ L+ \sum_{x=1}^{L} \ln ( \rho(x+1) w_{x+1}^- ) \right]
+  \sum_{x=1}^{L} \left[\rho(x) w_x^+ + \rho(x+1) w_{x+1}^-   \right] + O\left( \frac{1}{\vert j \vert}\right)
\label{rate2.5masterrhojd1jmoinsinfinity}
\end{eqnarray}
and 
\begin{eqnarray}
 I[\rho(x) ;j] 
&& \opsimeq_{j \to + \infty} 
L j \ln j - j \left[ L+ \sum_{x=1}^{L} \ln ( \rho(x) w_x^+ ) \right]
+  \sum_{x=1}^{L} \left[\rho(x) w_x^+ + \rho(x+1) w_{x+1}^-   \right] + O\left( \frac{1}{\vert j \vert}\right)
\label{rate2.5masterrhojd1jplusinfinity}
\end{eqnarray}
Eq. \ref{rate2.5masterrhojd1jplusinfinity}
is very similar to the Directed case of Eq. \ref{rate2.5directedinfinity}, since the backwards rates $w^-_x$ only
occur in the term of order $O(j^0)$ in the above expansion.

\subsection{ Large deviations for the density $\rho(x)$ alone  }

The optimization of Eq. \ref{rate2.5}
with respect to the current $j$
\begin{eqnarray}
0 && = \frac{ \partial I_{2.5}[\rho(x) ;j] }{ \partial j} 
= \sum_{x=1}^{L}
     \ln \left( \frac{\sqrt{ j^2 + 4 \rho(x) w_x^+ \rho(x+1) w_{x+1}^-  }  + j}{ 2\rho(x) w_x^+}  \right)  
\label{rate2.5deri}
\end{eqnarray}
yields that the optimal current $j_{opt}$ is the solution of
\begin{eqnarray}
1 && = \prod_{x=1}^{L} \left( \frac{\sqrt{ j_{opt}^2 + 4 \rho(x) w_x^+ \rho(x+1) w_{x+1}^-  }  + j_{opt}}{ 2\rho(x) w_x^+}  \right)
\label{rate2.5derijopt}
\end{eqnarray}
and is thus not as explicit as in the Directed case (Eq. \ref{joptdirected}),
when one wishes to compute the rate function for the density alone
\begin{eqnarray}
 I_{density}[\rho(x) ] = I[\rho(x) ;j_{opt}] 
=  \sum_{x=1}^{L}
&& \left[ \rho(x) w_x^+ + \rho(x+1) w_{x+1}^-  -\sqrt{ j_{opt}^2 + 4 \rho(x) w_x^+ \rho(x+1) w_{x+1}^-  }     \right]
\label{ri1rhomaster}
\end{eqnarray}

\subsection{ Large deviations for the current $j$ alone }

The optimisation of the rate function of Eq. \ref{rate2.5}
with respect to the density $\rho(x)$ submitted to the normalisation condition of Eq. \ref{normarho}
that can be taken into account via the Lagrange multiplier $k(j)$ that will depend on $j$
\begin{eqnarray}
0 && = \frac{ \partial}{ \partial \rho(x) } \left[   I[\rho(.) ;j] + k(j) \left( \sum_x \rho(x)-1 \right)  \right]
\label{rate2.5rhoderi}
\end{eqnarray}
is again not as explicit as for the Directed case of Eq. \ref{rate2.5masterrhojd1derirho} for arbitrary $j$.
We will thus focus on the limit of large currents $j \to \pm \infty$ 
where the rate function follows the expansion of 
 Eq. \ref{rate2.5masterrhojd1jmoinsinfinity} and
Eq. \ref{rate2.5masterrhojd1jplusinfinity},
so that the optimization of Eq. \ref{rate2.5rhoderi} then yields the optimal density
\begin{eqnarray}
\rho_{opt} (x)  && \opsimeq_{j \to \pm  \infty} \frac{ \vert j \vert }{ w_x^+ + w_x^- +k(j)   } 
\label{rhooptjinfty}
\end{eqnarray}
where the Lagrange multiplier $k(j)$ is fixed by the normalization of Eq. \ref{normarho}
\begin{eqnarray}
1 && = \sum_{x=1}^{L} \frac{ \vert j \vert }{ w_x^+ + w_x^- +k(j)   } 
\label{1rhooptjinfty}
\end{eqnarray}
so that for $j \to \pm \infty$, one obtains the asymptotic behavior of the Lagrange multiplier
\begin{eqnarray}
k(j) && \opsimeq_ {j \to \pm  \infty }  L j 
\label{kjinfty}
\end{eqnarray}
and the corresponding optimal uniform density
\begin{eqnarray}
\rho_{opt} (x)  && \opsimeq_{j \to \pm  \infty} \frac{ 1 }{L } 
\label{rhooptjinftyfin}
\end{eqnarray}
that one can plug into Eq. \ref{rate2.5masterrhojd1jmoinsinfinity}
and to obtain the asymptotic behavior of the rate function for the current alone 
\begin{eqnarray}
 I_{current}[j] =   I[\rho_{opt} (x) ;j] 
&& \opsimeq_{j \to - \infty} 
L \vert j \vert \ln \vert j \vert 
 + \vert j \vert \left[  L \ln L - L  - \sum_{x=1}^{L} \ln ( w_{x+1}^-  ) \right]
+ \frac{1}{L} \sum_{x=1}^{L} \left[w_x^+ +  w_{x+1}^-   \right] 
\label{ratejmoinsinfinity}
\end{eqnarray}
and
\begin{eqnarray}
I_{current}[j] =  I[\rho_{opt} (x) ;j] 
&& \opsimeq_{j \to + \infty} 
L j \ln j + j \left[  L \ln L - L - \sum_{x=1}^{L} \ln (  w_x^+ ) \right]
+\frac{1}{L} \sum_{x=1}^{L} \left[w_x^+ +  w_{x+1}^-   \right] 
\label{ratejplusinfinity}
\end{eqnarray}

\subsection{ Generating function $Z(\nu)$ of the current $j$ via the deformed Markov operator $W^{\nu}$  }

The deformed operator of Eq. \ref{wmasterdeformed} for the current alone 
reads 
\begin{eqnarray}
 W^{[\nu ]}  =
\sum_{x=1}^L \left( w_x^+ e^{\frac{\nu}{L}}  \vert x +1 \rangle \langle x \vert 
+ w_x^- e^{-\frac{\nu}{L}}  \vert x -1 \rangle \langle x \vert
 - (w_x^+ + w_x^-  ) \vert x  \rangle \langle x \vert \right)
\label{wmasterdeformednu}
\end{eqnarray}
so that the components of the right and left eigenvectors of Eq. \ref{wmastereigen}
now satisfy recurrence involving three consecutive components 
(instead of only two components in Eq. \ref{righteigen} and Eq. \ref{righteigen}  concerning the directed case).
\begin{eqnarray}
0 && = w_{x-1}^+ e^{\frac{\nu}{L}} R^{\nu}(x-1) + w_{x+1}^- e^{-\frac{\nu}{L}} R^{\nu}(x+1) - (w_x^+ + w_x^-  +E(\nu) ) R^{\nu}(x)
\nonumber \\
0 && = w_{x}^+ e^{\frac{\nu}{L}} L^{\nu}(x+1) + w_{x}^- e^{-\frac{\nu}{L}} L^{\nu}(x-1) - (w_x^+ + w_x^-  +E(\nu) ) L^{\nu}(x)
\label{eigensecond}
\end{eqnarray}
As a consequence, the relation between $\nu$ and $E(\nu)$ can be studied further
but is less explicit than for the Directed case (Eq. \ref{enueigen}).

However, the limits of large deformations $\nu \to \pm \infty$ of the Markov operator 
usually lead to simplifications and allow to determine the tails of the rate function of large currents $j \to \pm \infty$,
even in interacting models like exclusion processes (see the reviews \cite{lazarescu_companion,lazarescu_generic} and references therein).
In the following, we thus focus on these two limits $\nu \to \pm \infty$.

For $\nu \to +\infty$, corresponding to $j \to + \infty$, where the energy is also large $E(\nu) \to +\infty$,
the recurrences of Eq. \ref{eigensecond} can be approximated by the leading terms containing $ e^{\frac{\nu}{L}} $ and $E(\nu) $ 
\begin{eqnarray}
R^{\nu}(x) && \opsimeq_{\nu \to +\infty} \frac{ w_{x-1}^+ e^{\frac{\nu}{L}}}{ E(\nu)}  R^{\nu}(x-1)  
\opsimeq_{\nu \to +\infty}  \left[  \prod_{y=2}^x    \frac{ w_{y-1}^+ e^{\frac{\nu}{L}}}{ E(\nu)}   \right]R^{\nu}(1) 
\nonumber \\
  L^{\nu}(x) && \opsimeq_{\nu \to +\infty}  \frac{ E(\nu) } { w_{x-1}^+ e^{\frac{\nu}{L}} }  L^{\nu}(x-1)
\opsimeq_{\nu \to +\infty} \left[  \prod_{y=2}^x  \frac{ E(\nu) } { w_{y-1}^+ e^{\frac{\nu}{L}} }   \right]     L^{\nu}(1) 
\label{eigensecondvpinfty}
\end{eqnarray}
so that the periodic boundary conditions $R^{\nu}(L+1)=R^{\nu}(1)  $ and $L^{\nu}(L+1)=L^{\nu}(1)   $
yields that the eigenvalue $E(\nu)$ behave asymptotically as
\begin{eqnarray}
E(\nu) \opsimeq_{\nu \to +\infty} e^{ \displaystyle \frac{ \nu}{L}  +   \frac{1}{L} \sum_{x=1}^L \ln \left(   w^+_{x}  \right)    }
\label{nuelargebis}
\end{eqnarray}
i.e. exactly as in the Directed Model (Eq. \ref{nuelarge}).
Eq. \ref{nuelargebis} corresponds to the Legendre transform (Eq. \ref{legendrejnu}) of the leading terms of Eq. \ref{ratejplusinfinity} as it should.

For $\nu \to -\infty$, corresponding to $j \to - \infty$, the situation is of course completely different
from the Directed Trap model of the previous section where the current was constrained to be positive $j \geq 0$.
Here the energy is also large $E(\nu) \to +\infty$ in this limit $\nu \to -\infty $,
and the recurrences of Eq. \ref{eigensecond} can be approximated by the leading terms containing $ e^{-\frac{\nu}{L}} $ and $E(\nu) $ 
\begin{eqnarray}
 R^{\nu}(x) && \opsimeq_{\nu \to - \infty} \frac{ E(\nu)}{ w_{x}^- e^{-\frac{\nu}{L}}  }   R^{\nu}(x-1)
\opsimeq_{\nu \to -\infty}  \left[  \prod_{y=2}^x   \frac{ E(\nu)}{ w_{y}^- e^{-\frac{\nu}{L}}  }    \right]R^{\nu}(1) 
\nonumber \\
 L^{\nu}(x) && \opsimeq_{\nu \to - \infty}  \frac{ w_{x}^- e^{-\frac{\nu}{L}} } { E(\nu) } L^{\nu}(x-1) 
\opsimeq_{\nu \to -\infty} \left[  \prod_{y=2}^x  \frac{ w_{y}^- e^{-\frac{\nu}{L}} } { E(\nu) }   \right]     L^{\nu}(1) 
\label{eigensecondvminfty}
\end{eqnarray}
so that the periodic boundary conditions $R^{\nu}(L+1)=R^{\nu}(1)  $ and $L^{\nu}(L+1)=L^{\nu}(1)   $
yields that the eigenvalue $E(\nu)$ behave asymptotically as
\begin{eqnarray}
E(\nu) \opsimeq_{\nu \to -\infty} e^{ \displaystyle - \frac{ \nu}{L}  +   \frac{1}{L} \sum_{x=1}^L \ln \left(   w^-_{x}  \right)    }
\label{numoinsinfty}
\end{eqnarray}
that corresponds to the Legendre transform (Eq. \ref{legendrejnu}) of the leading terms of Eq. \ref{ratejmoinsinfinity} as it should.

As a final remark, let us mention that the fact that the optimal density becomes uniform for large currents $j \to \pm \infty$
 (as found in Eq. \ref{rhooptjinftyfin} via the direct contraction method) can be seen in the present deformed Markov operator approach
by considering the product $ L^{\nu}(x) R^{\nu}(x) $ of the components of the left and right eigenvectors
(see \cite{lazarescu_companion,lazarescu_generic,chetrite_conditioned} and references therein),
which happen indeed to be independent of $x$ in Eq. \ref{eigensecondvpinfty} and in Eq. \ref{eigensecondvminfty}.


\section{ Comparison with diffusion processes on the disordered ring }

\label{sec_diffusion}

In this last section, we briefly stress the differences with the properties of large deviations 
for diffusions in continuous space defined by Langevin equations.
To avoid the usual discussion between the Ito and Stratonovich conventions for Langevin stochastic differential equations,
let us define the dynamics directly via the Fokker-Planck Equation 
written as a continuity equation for the conservation of the probability $P_t(x)$
\begin{eqnarray}
\partial_t P_{t}(x)  && =- \partial_x j_t(x)
\nonumber \\
j_t(x) && = P_t(x) F(x) - D(x) \partial_x P_{t}(x) 
\label{fokkerplanck}
\end{eqnarray}
where the current $j_t(x)$ contains some random drift $F(x)$ and some random diffusion coefficient $D(x)$.
This formulation is then clearly the continuous analog of the Master Equation \ref{master}.

The steady state of Eq. \ref{fokkerplanck}
corresponds to a constant stationary current along the ring $ j_{st}(x)=j_{st}$,
and the stationary density $\rho_{st}(x)$ can be then explicitly computed in each disordered sample by solving
the differential equation
\begin{eqnarray}
j_{st} && = \rho_{st}(x)  F(x) - D(x) \partial_x  \rho_{st}(x) 
\label{fokkerplancksteady}
\end{eqnarray}
via the variation of constants, so that this stationary state is the continuous version of Eqs \ref{rhost} and Eq \ref{jst}.
The sum of products of random variables appearing in Eqs \ref{rhost} and Eq \ref{jst}
 correspond to the well-known structure of Kesten variables that show up in various discrete classical or quantum random models 
\cite{Kesten,Der_Pom,jpb_review,Der_Hil,Cal,c_microcano,us_watermelon,mblcayley,c_majtransfert},
Their continuous counterparts are known as exponential functionals of Brownian motion 
and have been also much studied \cite{flux,hyperbolic,yor,yor_book,texier,comtet_review} with very similar properties.

However, the large deviations for the joint distribution of the empirical density $\rho(x)$ and of the empirical current $j$
for the Fokker-Planck dynamics of Eq. \ref{fokkerplanck}
are described by the explicit rate function \cite{wynants_thesis,maes_diffusion,chetrite_formal,engel}
\begin{eqnarray}
&& I[ \rho(.),j]  = \frac{1}{4} \int_0^L \frac{dx}{D(x) \rho(x) } \left( j - \rho(x) F(x) +D(x) \frac{d \rho(x) }{dx}    \right)^2
\nonumber \\
&& = \frac{1}{4}
\left[ j^2  \int_0^L \frac{dx}{D(x) \rho(x) } 
-2 j  \int_0^L dx \left(  \frac{F(x)}{D(x)}  - \frac{d \ln (\rho(x) )}{dx}    \right) 
+ \int_0^L \frac{dx}{D(x) \rho(x) }   \left(  \rho(x) F(x) -D(x) \frac{d \rho(x) }{dx}    \right)^2
\right] 
\label{i2.5diffusion}
\end{eqnarray}
which is always Gaussian with respect to the current $j$,
in contrast to Eq. \ref{rate2.5masterrhojd1} for the discrete-space Master Equation of Eq. \ref{master}.
In particular, the Gaussian tails in $j^2$ for $j \to \pm \infty$ are completely different
from the tails of Eq. \ref{ratejmoinsinfinity} and Eq. \ref{ratejplusinfinity}.
This phenomenon is of course very general : the discrete-space formulation via Markov Jump processes and the continuous-space formulation
via diffusion processes may have similar behaviors in the small-fluctuations region near the typical state, but are completely different
in the large deviation region, where all details of the dynamics are relevant.
In particular, the tail in $(j \ln j)$ for the rate function $I_{current}(j)$ for large current $j \to +\infty$
is generic for Markov Jump processes without interactions as a consequence of the explicit general form of large deviations at Level 2.5
\cite{fortelle_jump,wynants_thesis,maes_canonical,maes_onandbeyond,chetrite_formal}
and is also present for Markov Jump processes with interactions like exclusions processes 
(see the reviews \cite{lazarescu_companion,lazarescu_generic} and references therein).
On the contrary, Gaussian tails in $j^2$ are generic for diffusion processes 
as a consequence of the explicit general form of large deviations at Level 2.5
\cite{wynants_thesis,maes_diffusion,chetrite_formal,engel}
or in the diffusive hydrodynamic approximation of interacting models like exclusion processes
known as 'Macroscopic Fluctuation Theory' (see the reviews  \cite{derrida-lecture,lazarescu_companion,lazarescu_generic}
and references therein).

\section{ Conclusions}

\label{sec_conclusion}

In this paper, we have analyzed the explicit form of the large deviations at level $2.5$ concerning the joint probability of the density $\rho(x)$ and of the current $j$ for the case of $N$ independent Markov Jump processes on a ring with random transition rates. 
We have first focused on the Directed Trap model, where the current is positive $j \in [0,+\infty[$.
The contractions needed to obtain the large deviations properties of the density alone and of the current alone give explicit results
in each disordered sample. In particular, with respect to the steady state density $\rho_{st}(x)$ whose inhomogeneity is fixed by the disorder,
the region of higher currents than in the steady state $j> j_{st}$ requires more homogeneous density $\rho(x)$ than $\rho_{st}(x)$,
while the region of smaller currents than in the steady state $j< j_{st}$ requires more inhomogeneous density $\rho(x)$ than $\rho_{st}(x)$.
We have also considered the deformed Markov operator needed to evaluate the generating function of the current 
in order to make a detailed comparison with the previous contraction approach.
We have then turned to the non-directed model where the current varies in $j \in ]-\infty,+\infty[$: while the contractions are not explicit for arbitrary values of the density or the current, we have shown how they can be solved to analyze the tails $j \to \pm \infty$.
Equivalently, the highest eigenvalue and the corresponding left and right eigenvectors of the deformed Markov operator 
can be explicitly studied for large deformations of the parameter $\nu \to \pm \infty$ conjugated to the current. 
Finally, we have stressed the differences with the Gaussian form of large deviations properties 
for diffusion processes on disordered rings.

\end{document}